\documentclass[aps,prl,twocolumn,showpacs,amsmath,amssymb,superscriptaddress]{revtex4}
\usepackage{graphicx}

\newcommand{\bem}{\begin{em}}
\newcommand{\eem}{\end{em}}
\newcommand{\eq}{\!=\!}
\newcommand{\bqn}{\begin{eqnarray*}}
\newcommand{\eqn}{\end{eqnarray*}}

\newcommand{\beq}{\begin{equation}}
\newcommand{\eeq}{\end{equation}}

\newcommand{\eps}{\epsilon}
\newcommand{\dt}{\delta t}

\renewcommand{\bar}[1]{\overline{#1}}

\begin{document}
  
\title{Thermodynamic Scaling of Diffusion in Supercooled Lennard-Jones Liquids}
\author{D.~Coslovich}
\affiliation{Dipartimento di Fisica Teorica, Universit{\`a} di Trieste
  -- Strada Costiera 11, 34100 Trieste, Italy} 
\affiliation{CNR-INFM Democritos National Simulation Center -- 
  Via Beirut 2-4, 34014 Trieste, Italy} 
\author{C.~M.~Roland}
\affiliation{Naval Research Laboratory, Code 6120, Washington, DC 20375-5342, USA} 
\date{\today}
\begin{abstract}
The manner in which the intermolecular potential $u(r)$ governs structural
relaxation in liquids is a long standing problem in condensed matter physics. 
Herein we show that diffusion coefficients for simulated Lennard-Jones $m$--6 liquids
($8\leq m \leq 36$) in normal and moderately supercooled states are a unique
function of the variable $\rho^{\gamma}/T$, where $\rho$ is density and $T$ is temperature. 
The scaling exponent $\gamma$ is a material specific constant whose
magnitude is related to the steepness of the repulsive part of
$u(r)$, evaluated around the distance of closest approach between
particles probed in the supercooled regime. Approximations of $u(r)$ in terms
of inverse power laws are also discussed. 
\end{abstract}


\maketitle

Establishing a quantitative connection between the relaxation properties of 
a liquid and the interactions among its constituent 
molecules is the \textit{sine qua non} for fundamental understanding and prediction of the 
dynamical properties. The supercooled regime is of particular interest, since 
both intermolecular forces and steric constraints (excluded volume) exert 
significant effects on the dynamics. This makes temperature, pressure, and 
volume essential experimental variables to characterize the relaxation 
properties. One successful approach to at least categorize dynamic properties
of supercooled liquids and polymers is by expressing them as 
a function of the ratio of mass density, $\rho$, to temperature, $T$, with the
former raised to a material specific constant $\gamma$,~\textit{viz.}  
\begin{equation}\label{eq1}
x = \Im(\rho^\gamma/T)
\end{equation}
where $x$ is the dynamic quantity under consideration, such as the structural
relaxation time $\tau$, the viscosity $\eta$, or the diffusion coefficient $D$, 
and $\Im$ is a function. 
This scaling was first applied to a
Lennard-Jones (LJ) fluid, with $\gamma=4$ yielding approximate master curves
of the reduced ``excess'' viscosity for different thermodynamic
conditions~\cite{ashurst75}. More recently Eq.~\eqref{eq1} has been shown to
superpose relaxation times measured by neutron scattering~\cite{tolle01}, light
scattering~\cite{dreyfus04}, viscosity~\cite{roland06}, and dielectric
spectroscopy~\cite{casalini04,albasimionesco04,urban05,reiser05,win06} for a
broad range of materials, including polymer blends and ionic liquids. 
The scaling exponent $\gamma $, which varies
in the range from 0.13 to 8.5~\cite{roland05}, is a 
measure of the contribution of density (or volume) to the dynamics, relative 
to that due to temperature. The only breakdown of the scaling is observed 
for hydrogen-bonded liquids, in which the concentration of H-bonds changes 
with $T$ and $P$, causing $\tau$ to deviate from
Eq.~\eqref{eq1}~\cite{roland06}. 

The function $\Im$ in Eq.~\eqref{eq1} is unknown \textit{a priori}. Its form
can be derived from entropy models for the glass transition, leading to an
exponential dependence of $\log{\tau}$ on
$\rho^{\gamma}/T$~\cite{casalini06,casalini07}. 
Another interpretation of the scaling is that the supercooled dynamics is
governed by activated processes with an effective activation energy
$E(\rho,T)$~\cite{tarjus04}, in which the
$\rho$-dependence of $E(\rho,T)$ can be factored and expressed in terms of a
power law of $\rho$. The power law scaling arose from the idea that the intermolecular 
potential for liquids can be approximated as a repulsive inverse power law
(IPL), with the weaker attractive forces treated as a spatially-uniform
background term~\cite{hoover71,chandler83,hm}
\begin{equation}\label{eq2}
u(r) \sim r^{-\bar{m}}+const
\end{equation}
where $r$ is the intermolecular distance. In the case of an IPL, in fact, all 
\bem reduced \eem dynamical quantities~\cite{hiwatari74} can be cast in the form
of Eq.~\eqref{eq1} with $\gamma=\bar{m}/3$, i.e., the thermodynamic scaling is
strictly obeyed. For instance, this applies to the reduced diffusion
coefficient $D^*\!\sim\!(\rho^{1/3}T^{-1/2})D \!\sim\!(T^{3/\bar{m}-1/2})D$. A
similar reduction of $D$ by macroscopic variables ($\rho$ and $T$) has also been
employed in entropy scaling laws of diffusion~\cite{rosenfeld99}. 

The IPL approximation emphasizes the dominant role of the short-range
repulsive interactions for local properties such as structural relaxation. 
Various groups have explored through numerical simulations the relationship of
the steepness of the repulsive potential to properties such as the equation of
state~\cite{lafitte06,heyes05,adamenko03}, longitudinal wave
transmission~\cite{anento90}, vibrational spectrum~\cite{kambayashi94},
liquid~\cite{galliero06} and gaseous~\cite{chapman70}
transport, the correlation between fluctuations of
energy and pressure~\cite{pedersen07}, and the 
fragility~\cite{bordat04,demichele04}. Recently two simulations have 
appeared in which Eq.~\eqref{eq1} was used to superpose dynamical data for
polymer chains described using an LJ $m$--6 potential with $m=12$ and an added
term for the intrachain interactions. The results appear contradictory: Tsolou
et 
al.~\cite{tsolou06} obtained a scaling exponent of $\gamma\!=\!2.8$ for the
segmental relaxation times of simulated 1,4-polybutadiene, while Budzien et
al.~\cite{budzien04} superposed diffusion coefficients for prototypical
polymer chains using $\gamma \eq 6$ when attraction were included in the
simulation and $\gamma \eq 12$ when they were omitted. Thus, the
scaling exponent $\gamma$ is either less than~\cite{tsolou06} or  
greater than~\cite{budzien04} $m/3$.

To clarify this situation and to establish a connection between the
thermodynamic scaling and the repulsive part of the intermolecular potential,
we carried out molecular dynamics simulations for supercooled LJ $m$--$n$
liquids, in which the repulsive exponent $m$ was systematically varied. 
Our models are binary mixtures composed of $N\!=\!500$
particles enclosed in a cubic box with periodic boundary conditions and
interacting with a LJ $m$--$n$ potential 
\beq\label{eqn:lj}
u_{\alpha\beta}(r) = 4 \epsilon_{\alpha\beta} \left[ {\left( 
    {\sigma_{\alpha\beta}}/{r} \right)}^{m} -
  {\left( 
    {\sigma_{\alpha\beta}}/{r} \right)}^{n} \right]
\eeq
where $\alpha,\beta\!=\!1,2$ are indexes of species. We fixed
the attractive exponent $n\!=\!6$, as in the standard LJ potential, and
varied $m\!=\!8,12,24,36$. The potential $u_{\alpha\beta}(r)$
was smoothed at $r_c\!=\!2.5\sigma_{\alpha\beta}$ using the cutoff scheme of
Stoddard and Ford~\cite{stoddard73}. Reduced LJ units are used, assuming
$\sigma_{11}$, $\epsilon_{11}$ and $\sqrt{m_1\sigma_{11}^2/\epsilon_{11}}$ as
units of distance, energy and time respectively. 
The mixture on which we focus is an additive, equimolar mixture with size ratio
$\lambda\!=\!\sigma_{22}/\sigma_{11}\!=\!0.64$, 
equal masses $m_1\!=\!m_2\!=\!1.0$ and a unique energy scale
$\eps_{\alpha\beta}\!=\!1.0$. The choice $m\!=\!12$ corresponds to the
AMLJ-0.64 mixture studied in~\cite{coslovich07a,coslovich07b}. The samples
were quenched isobarically at 
different pressures $P\!=\!5,10,20$ by coupling the system to Berendsen thermostat
and barostat during equilibration (see~\cite{coslovich07a} for
details), and performing the production runs in the NVE ensemble using the
Velocity-Verlet algorithm. The timestep $\dt$ was varied according to the
repulsive exponent, ranging from 0.001 ($m\!=\!36$) to 0.004 ($m\!=\!8$) at high $T$,
and from 0.003 ($m\!=\!36$) to 0.008 ($m\!=\!8$) at low $T$. The equilibration
criteria were similar to the ones used in~\cite{coslovich07a}.

\begin{figure}
\includegraphics*[width=0.47\textwidth]{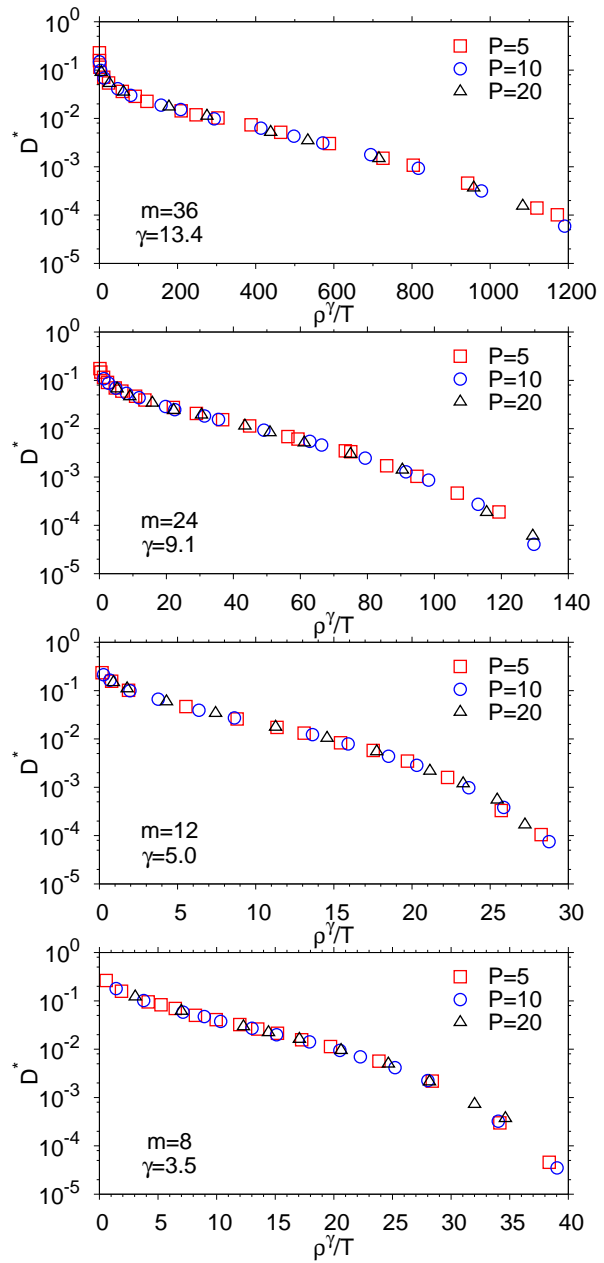}
\caption{\label{fig:scaling}
(color online). Reduced diffusion coefficients $D^*$ as a function of $\rho^\gamma/T$ for different
values of the repulsive exponent $m$ at different pressures: $P\!=\!5$
(squares), $P\!=\!10$ (circles), and $P\!=\!20$ (triangles). From top to
bottom: $m\!=\!36$ ($\gamma\!=\!13.4$), $m\!=\!24$ ($\gamma\!=\!9.1$),
$m\!=\!12$ ($\gamma\!=\!5.0$), and $m\!=\!8$ ($\gamma\!=\!3.5$). The estimated
uncertainty on $\gamma$ is $\pm 0.1$ ($\pm 0.2$ for $m=36$).}  
\end{figure}

The effectiveness of the thermodynamic scaling for LJ $m\!-\!6$ systems
is demonstrated in Fig.~\ref{fig:scaling} for different values of the
repulsive exponent $m$. 
For each $m$, reduced diffusion coefficients
$D^*\!=\!(\rho^{1/3}T^{-1/2})D$ were gathered along 
different isobaric paths ($P\!=\!5,10,20$) and the material specific scaling
exponent $\gamma$ was obtained by maximizing the overlap between different
sets of data, plotted as a function of $\rho^\gamma/T$.
Repeating the analysis for $D$, instead of $D^*$, yields very similar values
of $\gamma$, but the quality of the scaling for $D^*$ is slightly superior. 
Our data span roughly 5 decades of variation of $D$, 
over about two of which the temperature is lower than the so-called
onset temperature $T_O$~\cite{sastry98}, where non-exponential relaxation
typical of the supercooled regime first becomes apparent upon cooling the
liquid. 
Analyzing the variation of the scaling exponent in our models, 
we find that $\gamma$ increases with increasing $m$, but its actual 
value is systematically larger than $m/3$. For instance, in the case $m\!=\!12$ we 
obtain $\gamma\!=\!5.0$, a value which we also found to provide scaling of $D^*$ 
for other supercooled Lennard-Jones ($m\!=\!12$) mixtures, such as the AMLJ-0.76
mixture introduced in~\cite{coslovich07a} and the mixture of Kob and
Andersen~\cite{ka1}.

\begin{figure}
\includegraphics*[width=0.44\textwidth]{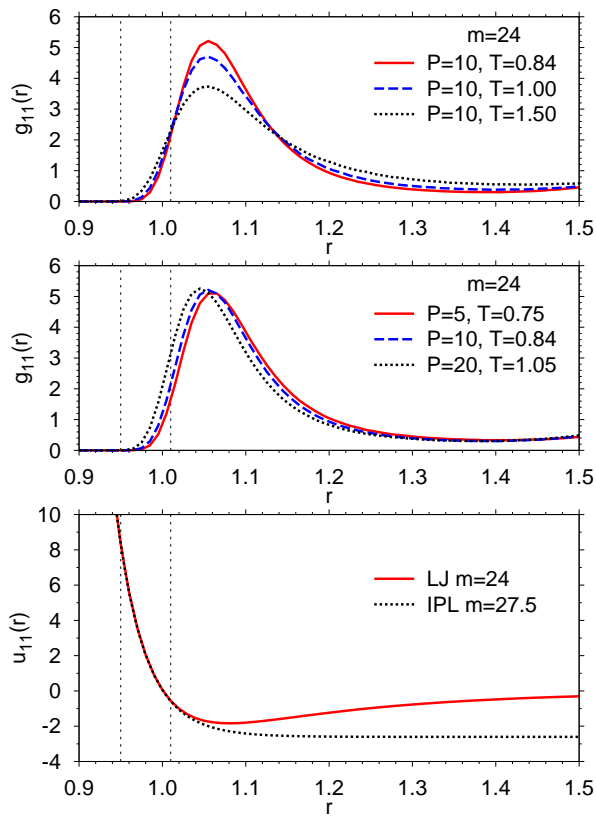}
\caption{\label{fig:repulsion}
(color online). Top panel: radial distribution functions between large particles $g_{11}(r)$
at $P=10$ for $T\alt T_O$: $T=1.20$
(dotted), $T\!=\!1.00$ (dashed), and $T=0.84$
(solid). Middle panel: $g_{11}(r)$ at the lowest equilibrated $T$: $T\!=\!0.75$ at $P\!=\!5$
(dotted), $T\!=\!0.84$ at $P\!=\!10$ (dashed), and $T\!=\!1.05$ at $P\!=\!20$
(solid). Bottom panel: pair potential $u_{11}(r)$ (solid) and fitted
IPL (dotted) in the range $[0.95:1.01]$. The latter range is indicated by 
vertical dotted lines in all panels.}   
\end{figure}

The origin of the discrepancy between $\gamma$ and $m/3$ lies in the fact that
the asymptotic region of small interparticle distances, in which $u(r)\sim
r^{-m}$, is not dynamically accessible in normal simulation conditions. The presence
of the fixed attractive term in the potential (Eq.~\eqref{eqn:lj}) gives rise
to an effective IPL which is steeper in the region of $r$ close to the minimum
than in the $r\rightarrow 0$ limit. 
This effect is illustrated in Fig.~\ref{fig:repulsion} for the case $m\!=\!24$.
The lower panel of Fig.~\ref{fig:repulsion} shows a fit of the pair
potential $u_{11}(r)$ to an IPL (Eq.~\eqref{eq2}) performed in the range
$[r_0:r_1]$, with $r_0\!=\!0.95$ and $r_1\!=\!1.01$. 
The value $\bar{m}\!=\!27.5$ obtained through
this procedure is indeed larger than $m\!=\!24$ and is in very good agreement
with the value expected from the dynamical scaling ($3\gamma\!=\!27.3\pm 0.03$). The range $[r_0:r_1]$
corresponds to typical distances of closest approach between particles
probed within our simulation conditions, as it can be seen by inspection
of the radial distribution functions $g_{11}(r)$ (see upper panels of
Fig.~\ref{fig:repulsion}).
Extending the range for the fit up to $r_1=1.06$, which is close to the
average position of the first peak in the $g_{11}(r)$, yields a larger value
$\bar{m}\!=\!28.8$, revealing how $\gamma$ is
dictated by the portion of $r$ around the distance of closest approach in the
supercooled regime.

\begin{table}
\begin{ruledtabular}
\caption{\label{table:exponents}
Parameters of IPL approximations for $u_{\alpha\beta}(r)$. The effective exponent
$\bar{m}$ is obtained from fitting to Eq.~\eqref{eqn:ipl}, whereas
$\bar{\eps}$, $\bar{k}$, and $\bar{x}$ are the optimal values for
Eq.~\eqref{eqn:ipl2}.}  
\begin{tabular}{rrrrrr}
          $m$ & $3\gamma$ & $\bar{m}$ & $\bar{x}$ & $\bar{\eps}$ & $\bar{k}$ \\    
\hline
          8  & 10.5(3) & 10.9  & 0.86 & 0.93 & $-1.05$ \\
          12 & 15.0(3) & 14.9  & 0.93 & 1.74 & $-1.80$ \\
          24 & 27.3(3) & 27.2  & 0.97 & 2.72 & $-2.74$ \\
          36 & 40.2(6) & 39.9  & 0.99 & 3.01 & $-3.01$ \\
\end{tabular}
\end{ruledtabular}
\end{table}

To proceed in a more systematic way, we considered all $\alpha\!-\!\beta$ pairs
(1-1, 1-2 and 2-2) in the potential $u_{\alpha\beta}(r)$ and performed a
simultaneous fit to the following IPL
\begin{equation}\label{eqn:ipl}
\bar{u}_{\alpha\beta}(r) = \bar{\eps} 
{\left( {\sigma_{\alpha\beta}}/{r} \right)}^{\bar{m}}
+ \bar{k} \, .
\end{equation}
The range for fitting was defined by two conventional
distances determined from the radial distribution functions
$g_{\alpha\beta}(r)$: the distance of 
closest approach between particles, $r_0$, (i.e., the value of $r$ for which 
the $g_{\alpha\beta}(r)$ first becomes non-zero) 
and the position corresponding to half of the height of the first peak, $r_1$,
(i.e., $g_{\alpha\beta}(r_1) = g_{\alpha\beta}(r_m)/2$ where $r_m$ is the
position of the first peak and $r_0\!<\!r_1\!<\!r_m$). 
These quantities depend on the thermodynamic 
state under consideration, but their variation with $P$ and $T$
is mild within our simulation conditions~\footnote{At fixed $P$,
$r_0$ and $r_1$ show a weak increase with decreasing $T$, but they become
almost $T$-independent below $T_O$.}. Our interest being the supercooled regime,
we simply consider the interval $[r_0:r_1]$ obtained from the low-$T$ 
behavior of the $g_{\alpha\beta}(r)$.
For each $\alpha\!-\!\beta$ pair we used the corresponding range $[r_0:r_1]$ for
fitting. In general, the fitted values of $\bar{m}$ are in good agreement with
$3\gamma$ (see Table~\ref{table:exponents}) for all values of $m$. Thus, the
scaling exponent can be reasonably accounted for in terms of an IPL
approximation of the pair potential, provided that a sensible choice of the
relevant range of distances is made. 

\begin{figure}
\includegraphics*[width=0.44\textwidth]{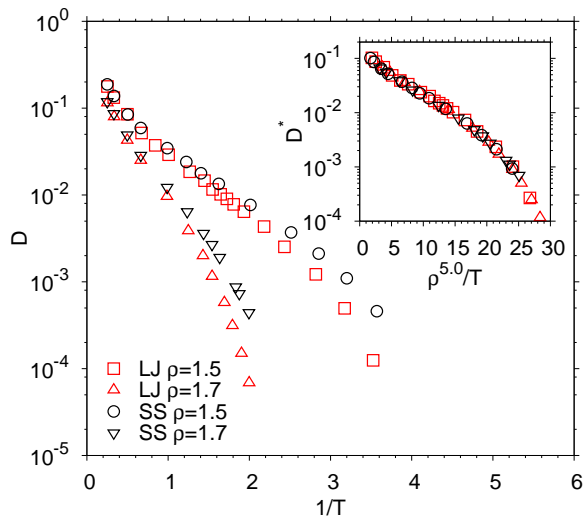}
\caption{\label{fig:ssreference}
(color online). Arrhenius plot of diffusion coefficient $D$ for the LJ 12--6 mixture and the
reference SS mixture ($\bar{m}=15.0$, $\bar{\eps}=1.74$) along two
isochores: $\rho=1.5$ and $\rho=1.7$. Inset: reduced diffusion coefficient
$D^*$ as a function of $\rho^{\bar{m}/3}/T$. For the SS mixture a reoptimized
energy scale $\tilde{\eps}=1.13\bar{\eps}$ was used.}   
\end{figure}

The above procedure suggests that a model of soft-spheres (SS) with
$\bar{m}=3\gamma$ should provide a good reference system for the LJ $m$--6
mixtures. To this aim, we approximate 
Eq.~\eqref{eqn:lj} with 
\begin{equation}\label{eqn:ipl2}
v_{\alpha\beta}(r) = 
\left\{
\begin{array}{ll}
\bar{\eps} {\left( {\sigma_{\alpha\beta}}/{r} \right)}^{\bar{m}} + \bar{k} & \quad r<\bar{x}\sigma_{\alpha\beta} \\
u_{\alpha\beta}(r)       & \quad r\geq \bar{x}\sigma_{\alpha\beta} \\
\end{array}
\right.
\end{equation}
where $\bar{m}$, $\bar{\eps}$, and $\bar{k}$ are expressed in terms of $\bar{x}$ by
requiring continuity of 0th, 1th, and 2th derivatives of $v_{\alpha\beta}(r)$ at
$r=x\sigma_{\alpha\beta}$. The value of $\bar{x}$ is then fixed by requiring that
$3\gamma=\bar{m}(\bar{x}) =
(m^2/\bar{x}^{m+1}-n^2/\bar{x}^{n+1})/(m/\bar{x}^{m+1}-n/\bar{x}^{n+1})$.   
The parameters defining the reference SS models for all values of $m$
are reported in Table~\ref{table:exponents}. We checked that the distance
$\bar{x}\sigma_{\alpha\beta}$ always lies in the range $[r_0:r_1]$ defined above.
Diffusivity data for the LJ 12--6 mixture are compared in
Fig.~\ref{fig:ssreference} to those of the 
corresponding reference SS mixture along two isochores ($\rho\eq 1.5$,
$\rho\eq 1.7$), which correspond to typical densities attained at low $T$ by the
LJ system (at constant $P$). The trend of $D(T)$ for the reference system
closely follows the one for the full LJ system. As expected, the SS mixture
has a larger diffusion coefficient for a given thermodynamic state. The
contribution to $D$ due to the attractive part of the potential could also be
explicitly included using a WCA-like splitting of
$v_{\alpha\beta}(r)$~\cite{straub92}.  
For the present purposes, however, it is more useful to note that a simple
rescaling of $\bar{\eps}$ (increased by around 10\%) yields an excellent
superposition of $D^*$ for all sets of data (see inset of
Fig.~\ref{fig:ssreference}). Thus, 
at least to a first approximation, the 
contribution of the attractive part of the potential to the dynamics alters
the shape of the function $\Im$ \bem without affecting \eem the scaling
exponent $\gamma$. 

To summarize, the thermodynamic scaling of the diffusion coefficient in
supercooled LJ $m$--6 mixtures reflects the importance of the repulsive part of
the pair potential in determining the dynamical properties of these
systems. The scaling exponent $\gamma$ is larger than $m/3$ for LJ $m$--6
liquids, a fact which can be rationalized by approximating the repulsive part
of the potential with an IPL having exponent $\bar{m}\approx 3\gamma$.
Generalizing such arguments to more realistic models of 
glass-formers~\cite{budzien04,tsolou06} and establishing connections with
other scaling procedures for $D^*$~\cite{dzugutov96,rosenfeld99} are open
challenges for future investigations.  

\begin{acknowledgments}
We thank G.~Pastore and J.~Dyre for useful discussions.  
Computational resources were obtained through a grant within the agreement
between the University of Trieste and CINECA
(Italy). The work at NRL was supported by the Office of Naval Research.
\end{acknowledgments}

\bibliographystyle{apsrev}

\end{document}